# Are Fluorination and Chlorination of the Morpholinium-Based Ionic Liquids Favorable?

Vitaly V. Chaban[1] and Oleg V. Prezhdo[2]

[1] Instituto de Ciência e Tecnologia, Universidade Federal de São Paulo, 12231-280, São José dos Campos, SP, Brazil

[2] Department of Chemistry, University of Southern California, Los Angeles, CA 90089, United States

**Abstract**. Room-temperature ionic liquids (RTILs) constitute a fine-tunable class of compounds. Morpholinium-based cations are new to the field. They are promising candidates for electrochemistry, micellization and catalytic applications. We investigate halogenation (fluorination and chlorination) of the N-ethyl-N-methylmorpholinium cation from thermodynamics perspective. We find that substitutional fluorination is much more energetically favorable than substitutional chlorination, although the latter is also a permitted process. Although all halogenation at different locations are possible, they are not equally favorable. Furthermore, the trends are not identical in the case of fluorination and chlorination. We link the thermodynamic observables to electron density distribution within the investigated cation. The reported insights are based on the coupled-cluster technique, which is a highly accurate and reliable electron-correlation method. Novel derivatives of the morpholinium-based RTILs are discussed, motivating further efforts in synthetic chemistry.

**Key words**: ionic liquids; morpholinium; fluorine; chlorine; thermodynamics.

TOC Graphic

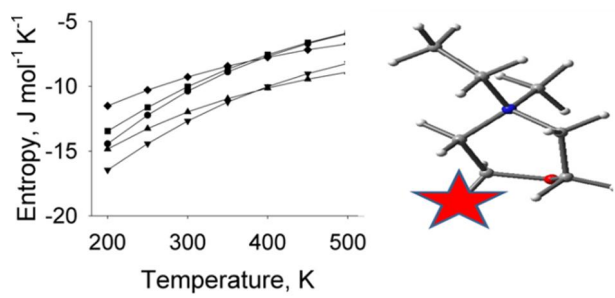

**Introduction**

Room-temperature ionic liquids (RTILs) are actively investigated in various branches of pure and applied science.[1-4] They are employed as universal solvents, electrolytes, catalysts, and reaction media. RTILs are used in separation applications, fuel and solar cells, gas capture and even biotechnology.[1-19] Due to their specific structural properties, RTILs cannot form an energetically favorable crystalline lattice. Consequently, they exist in the liquid state over an unusually wide temperature range. A single cation can be coupled with numerous anions providing a universe of novel compounds with tunable physical-chemical properties.

Most RTILs are based on heterocyclic organic molecules, such as imidazole, pyrrolidine, pyridine, piperidine, etc. Both protic and aprotic RTILs can be synthesized in many cases. Protons belonging to cations play an important role in formation of cation-anionic and cation-molecular hydrogen bonds. Hydrogen bonds are able to adjust physical properties of the solvent drastically. Morpholinium-based RTILs[20-23] have been characterized recently. They have obtained significant attention due to their specific structural properties.[23] Morpholinium-based cations are bipolar owing to simultaneous presence of oxygen (electron-rich center) and nitrogen (electron-poor center). They can be useful to design ionic liquid crystals. The crystals of the morpholinium-based compounds with sulfosuccinate anions and long alkane chains, $C_8$ to $C_{18}$, exhibit hexagonal columnar phases at room temperature.[24]

Morpholinium-based RTILs are practical in electrolyte applications due to straightforward synthesis, low cost, and high mobility of the lithium cation.[25] Lithium interacts strongly with the oxygen group in the morpholinium-based cation. Such cation-cation interactions draw strong interest in view of electrolyte applications, such as in batteries, solar cells, and supercapacitors. Unfortunately, these RTILs are viscous. They reveal a large electrochemical window in conjunction with reasonable conductivity. According to Galinski and Stepniak, mixtures of the morpholinium-based RTILs with propylene carbonate are prospective electrolytes for electrochemical double-layer capacitors.[26] These authors reported temperature dependence of ionic conductivities and electrochemical stabilities of N-ethyl-N-butylmorpholinium and N-butyl-N-methylmorpholinium bis(trifluoromechanesulfonyl)imides in mixtures with propylene carbonate. It was found that the conductivity maximum for these mixtures is located at 20 mol% of RTIL.

An ability of the morpholinium cation to participate in micellization was recently reported.[20] The alkyl chain is responsible for an interplay between hydrophobic and hydrophilic interactions determining micellization behavior. Besides that, morpholinium-based RTILs are interesting as prospective catalysts for organic synthesis, heat stabilizers, oxidants and corrosion inhibitors. Morpholinium-based cations were coupled with tetrafluoroborate, hexafluorophosphate, bis(trifluoromethanesulfonyl)imide, formate, bromide, as well as with anions decorated with long alkyl chains.

Domanska and Lukoshko[27] measured activity coefficients at infinite dilution for N-butyl-N-methylmorpholinium tricyanomethanide in 61 solvents, including alkanes, cycloalkanes, alkenes, alkynes, aromatic hydrocarbons, alcohols, water, ethers, ketones, acetonitrile. As compared to N-butyl-N-methylpyrrolidinium tricyanomethanide, these RTILs exhibits higher selectivity in separation of aromatic hydrocarbons from aliphatic hydrocarbons. Successful separation of thiophene and pyridine from heptane was highlighted. The demonstrated selectivity towards sulfur and nitrogen containing compounds is interesting for many prospective applications of ionic liquids.

Halogenation of the morpholinium-based cations has never been reported before. This work investigates fluorination and chlorination of the N-ethyl-N-methylmorpholinium cation from the thermodynamic perspective. We identify five chemically non-equivalent prospective halogenation sites in the N-ethyl-N-methylmorpholinium cation. Not only halogenation of the side hydrocarbon chains (ethyl and methyl) was considered, but also halogenation of the morpholine ring was investigated. We show that substitutional fluorination is thermodynamically more favorable in all cases. Chlorination is also permitted, but it is ca. 4 times less favorable. These insights were obtained using highly accurate ab initio coupled-cluster electronic structure calculations.[28] The resulting molecular partition functions were further processed to obtain thermodynamic quantities, such as enthalpy, entropy, free energy of reaction and their temperature dependence.

**Methodology**

Fluorination and chlorination of the N-ethyl-N-methylmorpholinium cation is investigated from the thermodynamic perspective. The direction of a chemical reaction taking part at constant temperature and constant pressure is determined by evolution of the corresponding Gibbs free energy, $G = f(T,P)$. Gibbs free energy is expressed through a linear combination of enthalpy (H) and entropy (S) as $G = H - TS$, where T stands for temperature in Kelvin. We report these functions for the halogenation reactions and their temperature dependence over the temperature range between 200 and 500 K.

The internal energies, enthalpies, entropies and free energies are derived from the molecular partition function using the equations of statistical mechanics. The partition function is derived from a quantum-chemical calculation and subsequent frequency analysis for a given chemical entity. All thermodynamics quantities computed in this way correspond to an ideal gas. It is important that an electronic structure method used for wave-function optimization accounts for electron correlation. This is because major numerical errors arise from lack of electron correlation, even though a reliable equilibrium geometry of the molecule can be obtained at the Hartree-Fock or density functional theory levels.

The electronic structure of all systems was optimized using the coupled-cluster technique.[28] Coupled-cluster was born in nuclear physics, but received its recognition years later when applied to many-body electronic systems.[28] The coupled-cluster technique belongs to the group of post-Hartree-Fock methods. That is, any calculation using the coupled-cluster technique starts with a conventional one-electron calculation, and then the correlation between electrons is accounted for. This method provides very accurate energies of electron levels. Unfortunately, the underlying computational cost is high, and consideration of large molecules is thus prohibited. The implementation of coupled-cluster employed in the present work uses single and double substitutions from the Hartree-Fock determinant. Furthermore, it includes triple excitations non-iteratively.[29]

The split-valence triple-zeta 6-311G Pople basis set with added polarization and diffuse functions was applied. All electrons were considered explicitly. The pristine cation contains 72 electrons, the fluorinated cation contains 80 electrons, and the chlorinated cation contains 88 electrons. The wave function convergence criterion at every self-consistent field (SCF) step was set to $10^{-8}$ Hartree. The described electronic structure methods are available in the GAMESS simulation suite.[30]

**Results and Discussion**

We selected N-ethyl-N-methylmorpholinium as a prototypical cation. The shorter hydrocarbon chain is methyl, while the longer chain is ethyl, as it is in the currently available experiments. The terminating methyl group of ethyl is weakly influenced by the morpholine ring. This is also the case in imidazolium- and pyridinium-based RTILs.[12,31,32] Indeed, this group appears nearly neutral. For this reason, longer hydrocarbon chains can be avoided in simulaion, in order to save computational resources. Since morpholinium-based cations are rather new to chemists, widely applied nomenclature of their interaction sites is absent. We propose to apply the following nomenclature to label the symmetrically non-equivalent carbon atoms (Figure 1). C(N) denotes the two carbon atoms linked to nitrogen, while C(O) denotes the two carbon atoms linked to oxygen. Recall that nitrogen is the most electron deficient interaction site in this cation, while oxygen is the most electron rich site. The carbon atoms in the side chains are named as usual: CH$_3$ for methyl group; CH$_2$ for methylene group; ET for the terminating methyl group in the ethyl chain.

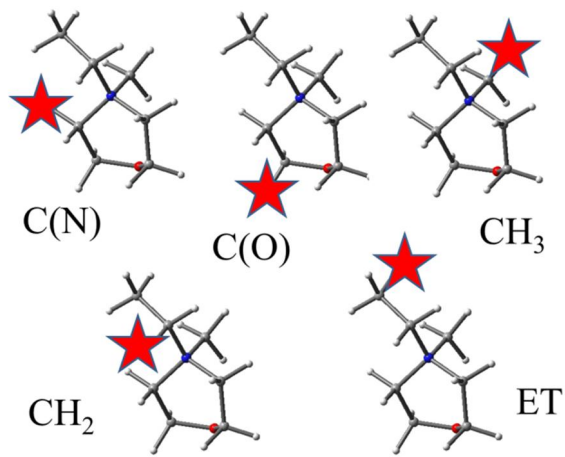

**Figure 1**. Halogen substitution sites in the N-ethyl-N-methylmorpholinium cation. Carbon atoms are grey, hydrogen atoms are white, nitrogen atom is blue, oxygen atom is red, and the halogen (fluorine, chlorine) atom is marked with a red star. As it was found by quantum-chemical calculations, five symmetrically non-equivalent carbon atoms are available.

The morpholinium-based cation is a significantly different chemical entity, compared to many other previously known RTIL cations, such as imidazolium, pyridinium, piperdidinium, pyrrolidinium, tetraalkylammonium, cholinium, etc. All the enumerated cations are monopolar, since they contain only a positively charged interaction center. This can be a single center such as the nitrogen atom in the pyridine ring or two centers such as the two nitrogen atoms in the imidazole ring. In turn, the N-ethyl-N-methylmorpholinium cation contains one positively charged center (nitrogen) and one negatively charged center (oxygen), whereas all other atoms exhibit smaller charges. The electrostatic charge on the nitrogen atom amounts to +0.29e. It almost compensates the opposite charge of the oxygen atom, which is equal to -0.37e. The two carbon atoms located near oxygen have +0.19e, and the two carbon atoms near nitrogen have - 0.19e and -0.10e. The symmetry near nitrogen is broken due to the strong influence of the methyl group, which is electron deficient, +0.16e. Another difference of the morpholinium cation from most other RTIL cations is that its ring is clearly non-planar. For instance, the O-C-C-N angle equals to 32 degrees. The C-O-C angle is 112 degrees and the C-N-C angle is 108 degrees. Compare these values to an ideal angle of 120 degrees in the planar hexagon, such as benzene. The observations regarding geometry and electron density distribution in the N-ethyl-N-methylmorpholinium cation are important for understanding of the halogenation trends discussed below.

Figure 2 summarizes the thermodynamic potentials for the fluorination reaction taking place via the five symmetrically different sites: C(N), C(O), $CH_3$, $CH_2$, and ET. Enthalpy is the decisive factor determining the reaction direction. The difference in enthalpy between the products ($C_7H_{15}NOF$ and HF) and the reactants ($C_7H_{16}NO$ and $F_2$) is significantly negative, indicating that carbon-fluorine covalent bonds in the morpholine ring are more stable than carbon-hydrogen bonds. Favorable halogenation of the hydrocarbon chains was rather expected from conventional chemical wisdom. Isobaric heat capacities are small. The reactions become slightly less favorable as temperature increases. Entropy changes are negative, that is, the entropy factor prohibits halogenation. However, the calculated entropy factor is too small compared to the enthalpy changes.

The trend in free energy completely repeats the trend in enthalpy. The most favorable fluorination site is C(O). This is not easy to predict without an appropriate numerical analysis, since one would rather expect straightforward fluorination of the hydrocarbon chains. It is well-known that hydrocarbons are readily fluorinated in the laboratory synthesis. The easier fluorination via the C(O) site indicates a strong role of the oxygen atom, which pulls electronic density of the adjacent carbon atoms towards itself. Having attained partially positive charges, the C(O) carbon atoms react with fluorine, which is strongly electronegative. More polar covalent bonds are formed. The same reasoning applies to the methylene group site, which is ca. 10 kJ mol$^{-1}$ less favorable. The differences between the remaining interaction sites are smaller, although they still exceeds kT.

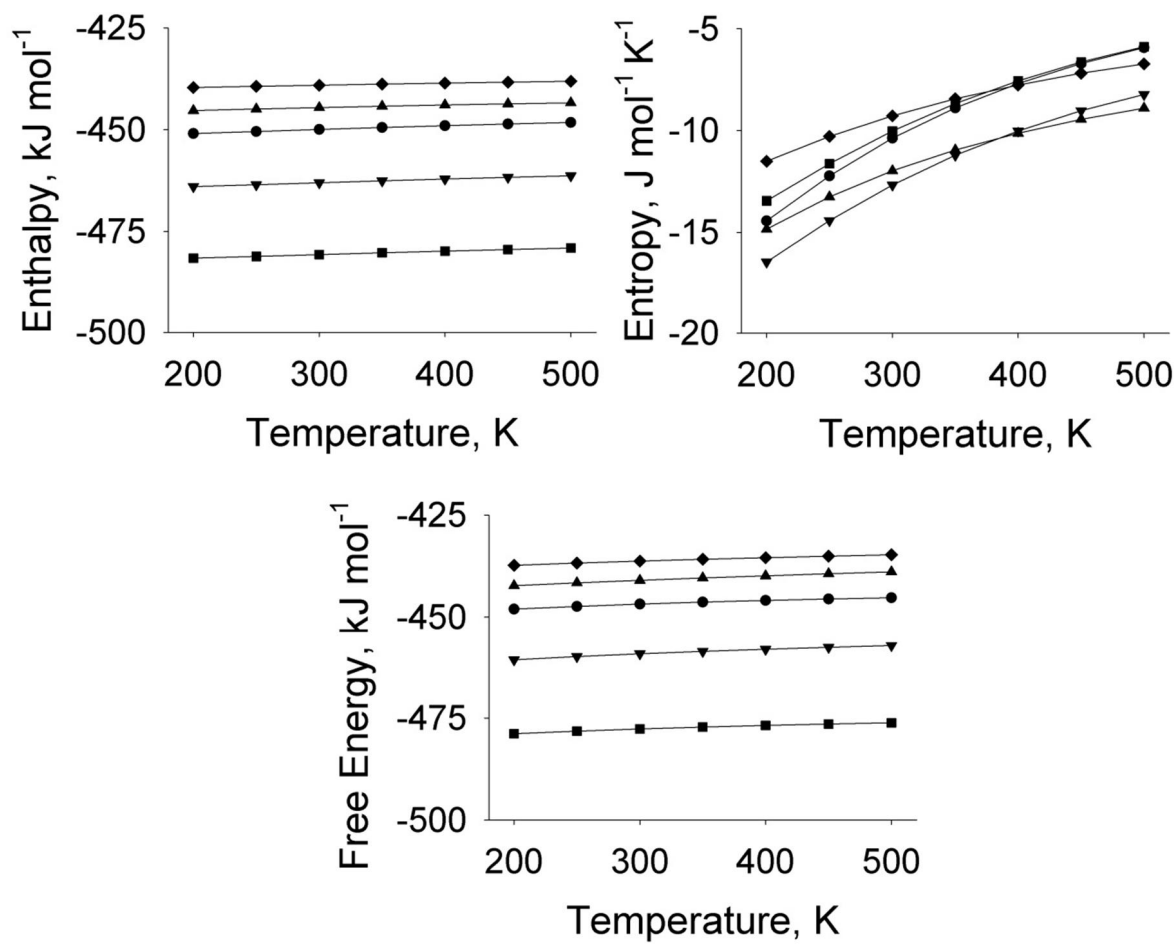

**Figure 2**. Enthalpy, entropy and free energy change during the substitutional fluorination reaction of the N-ethyl-N-methylmorpholinium cation, which takes place via the C(N) (circles), C(O) (squares), CH$_3$ (triangles up), CH$_2$ (triangles down), and ET (diamonds) reaction sites, see Figure 1.

Chlorination also appears favorable for all the perspective reaction sites. However, the free energy gain (Figure 3) is ca. four times smaller than that in the case of fluorination. This may be due to weaker carbon-chlorine bonds and also, to a certain extent, due to steric hindrances. The chlorine atom is significantly larger than the fluorine atom. Similar to fluorination, the C(O) site is most easily chlorinated, and the terminal atom of the ethyl chain is relatively resistant to this reaction. Chlorination of the methyl group is easier than chlorination of the methylene group, although it was contrariwise in the case of fluorination. It is probable that due to its size, the chlorine atom perturbs geometry of this part of the N-ethyl-N-methylmorpholinium cation.

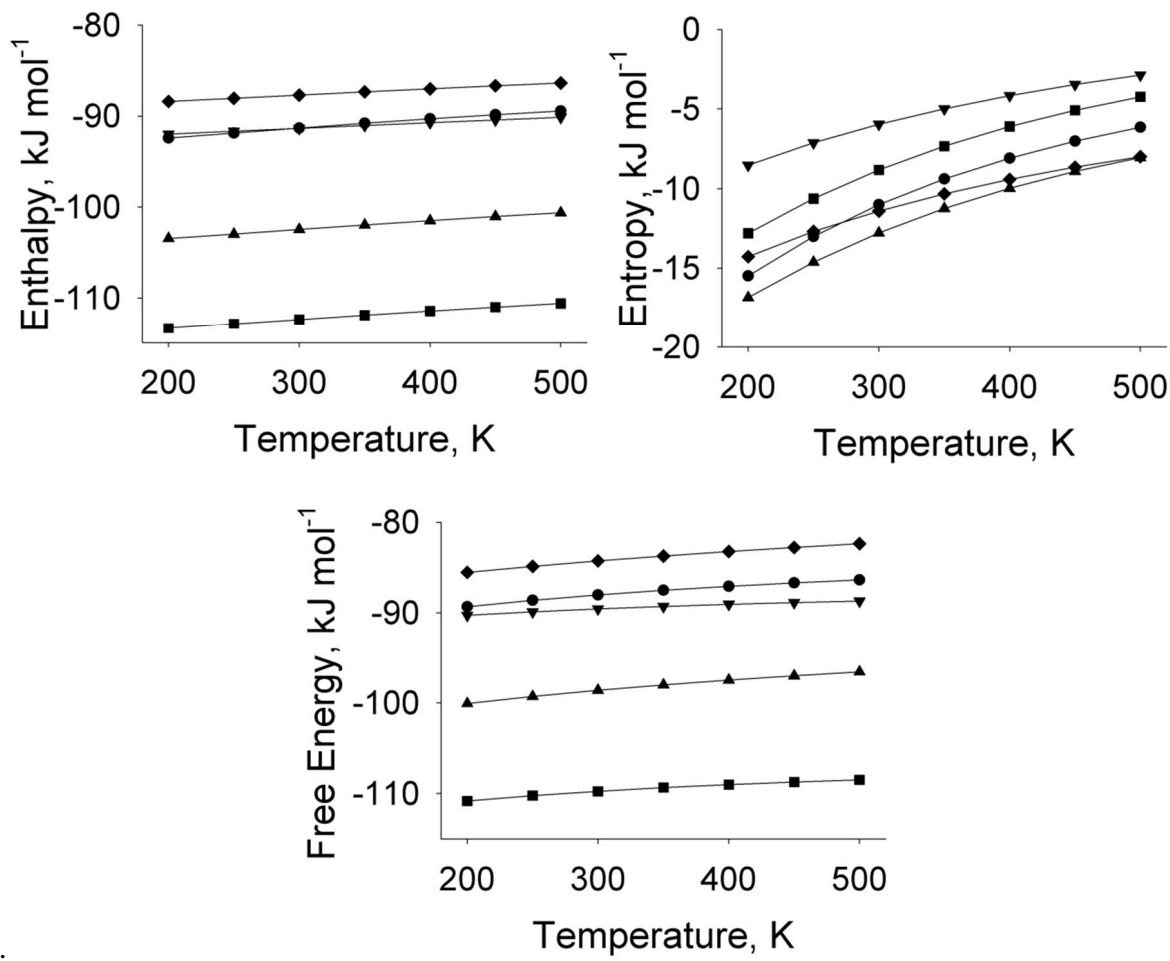

.

**Figure 3**. Enthalpy, entropy and free energy change during the substitutional chlorination reaction of the N-ethyl-N-methylmorpholinium cation, which takes place via the C(N) (circles), C(O) (squares), CH$_3$ (triangles up), CH$_2$ (triangles down), and ET (diamonds) reaction sites, see Figure 1.

Analysis of electrostatic charges on the fluorine and chlorine atoms (Figures 4-5) is helpful in understanding the thermodynamic observations and trends. The fluorine atom is negative in all positions of the reaction product. This is expected due to a high electron affinity of this element. However, prediction of particular values was challenging. As discussed above, the C(O) site indeed favors the largest negative electronic charge of fluorine due to proximity to another electronegative atom, oxygen. The terminal methyl group of the ethyl chain is close to neutrality, q=+0.05e, in the pristine morpholinium cation. The fluorine atom makes it electron deficient, q=+0.13e, by inducing an unusually large positive charge on the carbon atom, q=+0.32e. This makes fluorination reaction at this site less favorable, according to Figure 2.

The chlorine atom can be both positively and negatively charged, although these charges are small. Note that a positive charge on chlorine in the morpholinium cation is favored by the fact that the overall structure must maintain the +1e charge. If two atoms with different electronegativities are bonded, it does not implicitly mean that one of them will be necessarily electron deficient and another one will be electron rich. Both can be positively charged (with different charges according to electronegativity) depending on their particular position in the ion structure.

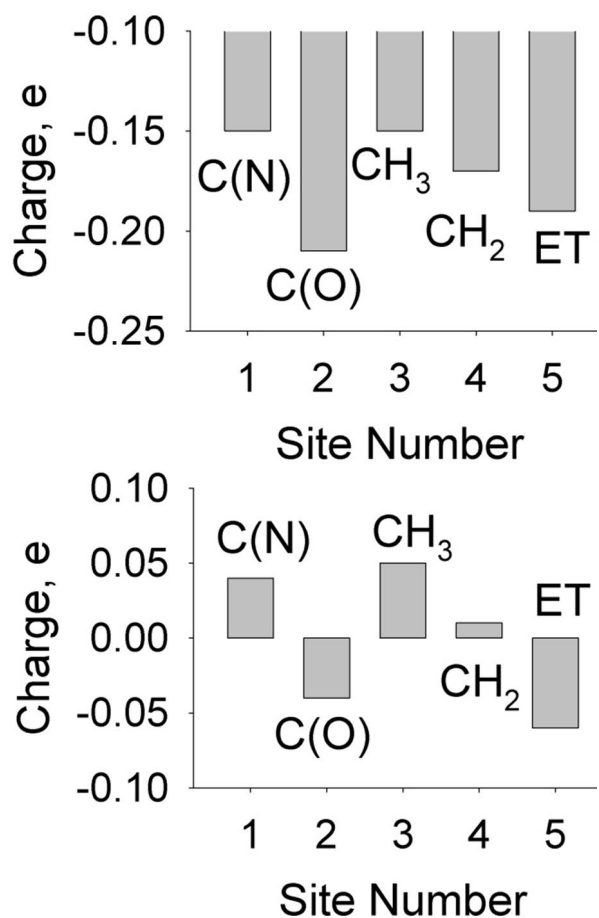

**Figure 4**. Electrostatic potential fitted charge on the fluorine (top) and chlorine (bottom) atoms in the N-ethyl-N-methylmorpholinium cation in the five investigated positions: C(N), C(O), CH$_3$, CH$_2$, and ET, see Figure 1.

Mulliken charges are derived directly from the optimized wave function. They represent a convenient measure of the electron population at a given atom, although their absolute values are not recommended as a characteristic of electrostatic interactions. It is interesting to note that chlorine in the methyl group, CH$_2$Cl, has a small, but positive charge. The chlorination at CH$_3$ is relatively favorable being inferior only to the chlorination at the C(O) site. In turn, fluorination at CH$_3$ is less favorable, likely because fluorine cannot acquire a positive charge. The methyl group is electron deficient, q=+0.21e. This partial charge is important to maintain the total +1e charge on the cation. Attachment of fluorine necessitates redistribution of the electron density over the entire N-ethyl-N-methylmorpholinium cation.

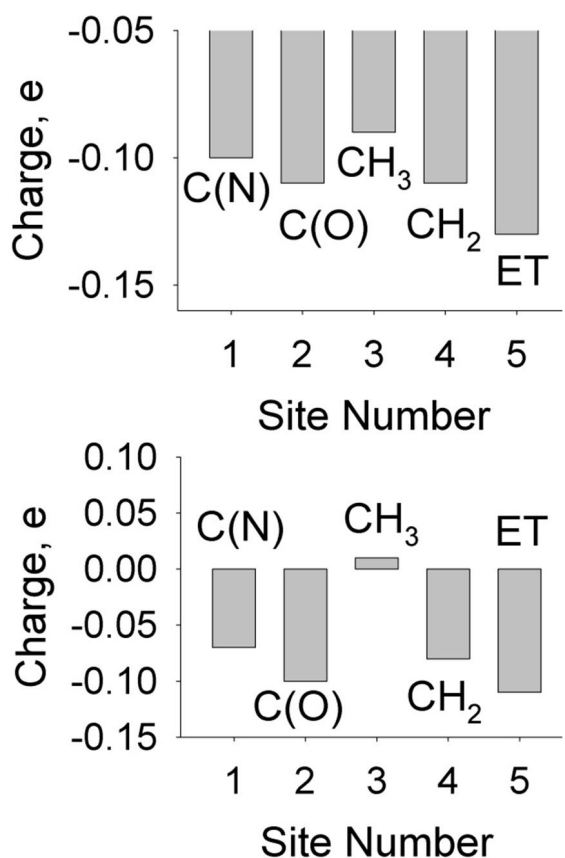

**Figure 5**. Mulliken charge on the fluorine (top) and chlorine (bottom) atoms in the morpholinium cation in the five investigated positions: C(N), C(O), CH$_3$, CH$_2$, and ET, see Figure 1.

**Concluding Remarks**

The present work, for the first time, reports a theoretical study of the fluorination and chlorination reactions of the prototypical morpholinium cation used in RTILs. Thermodynamic potentials of all reactions are discussed in relationship to the electronic structure and geometry of the cation. We show that halogenation of the cation is energetically favorable in all considered cases. Most likely, the described ionic derivatives will be soon synthesized starting from the morpholinium-based cations and using the well-established methods for fluorination (e.g. treatment by potassium fluoride) and chlorination (e.g. treatment by gaseous chlorine). Fluorination brings a larger free energy gain.

Substitution of hydrogen atoms by halogen atoms is an important tool in tuning physical-chemical properties of ionic liquids. Unlike hydrogen, fluorine and chlorine are unable to maintain hydrogen bonding with the counterion and polar solvent molecules, if any are present. Hydrogen bonding is an important phenomenon determining identity of many RTILs. Its presence is favorable for certain applications, but it is unfavorable for other applications. Furthermore, fluorination is expected to lead to more polar bonds, i.e. higher partial charges at the corresponding interaction centers. Halogenation is an important tool in tuning hydrophobicity/hydrophilicity balance of RTILs. Understanding of halogenation regularities is necessary for future progress in the field of RTILs, while relatively little has been performed in this direction thus far.

It must be noted that our results were obtained under an ideal gas approximation. That is, no ion-molecular interactions were considered, no solvation effects and no steric hindrances were accounted for. Furthermore, excited states are not considered. This should be a valid approximation for ionic liquids, in which HOMO-LUMO gaps are significant.

The reported simulations of chemical thermodynamics encourage experimental efforts in organic chemistry and provide guide for further development of RTILs. The fluorinated and chlorinated morpholinium-based cations are expected to exhibit significantly different physical chemical properties than the original cations.

**Acknowledgments.** This project was partially funded by CAPES (Brazil). OVP acknowledges financial support of the U.S. National Science Foundation, grant CHE-1300118.

**Contact Information.** E-mail for correspondence: vvchaban@gmail.com. Tel: +55 12 3309-9573; Fax: +55 12 3921-8857.

**REFERENCES**

(1) Chaban, V. V.; Prezhdo, O. V. Ionic and Molecular Liquids: Working Together for Robust Engineering. *Journal of Physical Chemistry Letters* 2013, *4*, 1423-1431.
(2) Patel, R.; Kumari, M.; Khan, A. B. Recent Advances in the Applications of Ionic Liquids in Protein Stability and Activity: A Review. *Applied Biochemistry and Biotechnology* 2014, *172*, 3701-3720.
(3) Guo, F.; Zhang, S. J.; Wang, J. J.; Teng, B. T.; Zhang, T. Y.; Fan, M. H. Synthesis and Applications of Ionic Liquids in Clean Energy and Environment: A Review. *Current Organic Chemistry* 2014, *19*, 455-468.
(4) Pandey, S. Analytical Applications of Room-Temperature Ionic Liquids: A Review of Recent Efforts. *Analytica Chimica Acta* 2006, *556*, 38-45.
(5) Kaintz, A.; Baker, G.; Benesi, A.; Maroncelli, M. Solute Diffusion in Ionic Liquids, Nmr Measurements and Comparisons to Conventional Solvents. *Journal of Physical Chemistry B* 2013, *117*, 11697-11708.
(6) Chaban, V. Hydrogen Fluoride Capture by Imidazolium Acetate Ionic Liquid. *Chemical Physics Letters* 2015, *625*, 110-115.
(7) Bica, K.; Deetlefs, M.; Schroder, C.; Seddon, K. R. Polarisabilities of Alkylimidazolium Ionic Liquids. *Physical Chemistry Chemical Physics* 2013, *15*, 2703-2711.
(8) Emel'yanenko, V. N.; Zaitsau, D. H.; Verevkin, S. P.; Heintz, A.; Voss, K.; Schulz, A. Vaporization and Formation Enthalpies of 1-Alkyl-3-Methylimidazolium Tricyanomethanides. *Journal of Physical Chemistry B* 2011, *115*, 11712-11717.
(9) Chaban, V. V.; Prezhdo, O. V. How Toxic Are Ionic Liquid/Acetonitrile Mixtures? *Journal of Physical Chemistry Letters* 2011, *2*, 2499-2503.
(10) Xu, C. H.; Margulis, C. J. Solvation of an Excess Electron in Pyrrolidinium Dicyanamide Based Ionic Liquids. *Journal of Physical Chemistry B* 2015, *119*, 532-542.
(11) Li, H.; Endres, F.; Atkin, R. Effect of Alkyl Chain Length and Anion Species on the Interfacial Nanostructure of Ionic Liquids at the Au(111)-Ionic Liquid Interface as a Function of Potential. *Physical Chemistry Chemical Physics* 2013, *15*, 14624-14633.
(12) Chaban, V. V.; Voroshyloya, I. V.; Kalugin, O. N.; Prezhdo, O. V. Acetonitrile Boosts Conductivity of Imidazolium Ionic Liquids. *Journal of Physical Chemistry B* 2012, *116*, 7719-7727.
(13) Niedermeyer, H.; Ab Rani, M. A.; Lickiss, P. D.; Hallett, J. P.; Welton, T.; White, A. J. P.; Hunt, P. A. Understanding Siloxane Functionalised Ionic Liquids. *Physical Chemistry Chemical Physics* 2010, *12*, 2018-2029.
(14) Castner, E. W.; Margulis, C. J.; Maroncelli, M.; Wishart, J. F. Ionic Liquids: Structure and Photochemical Reactions. *Annual Review of Physical Chemistry, Vol 62* 2011, *62*, 85-105.
(15) Keskin, S.; Kayrak-Talay, D.; Akman, U.; Hortacsu, O. A Review of Ionic Liquids Towards Supercritical Fluid Applications. *Journal of Supercritical Fluids* 2007, *43*, 150-180.
(16) Silvester, D. S.; Compton, R. G. Electrochemistry in Room Temperature Ionic Liquids: A Review and Some Possible Applications. *Zeitschrift Fur Physikalische Chemie-International Journal of Research in Physical Chemistry & Chemical Physics* 2006, *220*, 1247-1274.
(17) Shamsuri, A. A.; Daik, R. Applications of Ionic Liquids and Their Mixtures for Preparation of Advanced Polymer Blends and Composites: A Short Review. *Reviews on Advanced Materials Science* 2015, *40*, 45-59.
(18) Fileti, E. E.; Chaban, V. V. Imidazolium Ionic Liquid Helps to Disperse Fullerenes in Water. *Journal of Physical Chemistry Letters* 2014, *5*, 1795-1800.
(19) Hantal, G.; Voroshylova, I.; Cordeiro, M. N. D. S.; Jorge, M. A Systematic Molecular Simulation Study of Ionic Liquid Surfaces Using Intrinsic Analysis Methods. *Physical Chemistry Chemical Physics* 2012, *14*, 5200-5213.
(20) Kamboj, R.; Bharmoria, P.; Chauhan, V.; Singh, S.; Kumar, A.; Mithu, V. S.; Kang, T. S. Micellization Behavior of Morpholinium-Based Amide-Functionalized Ionic Liquids in Aqueous Media. *Langmuir* 2014, *30*, 9920-9930.
(21) Brigouleix, C.; Anouti, M.; Jacquemin, J.; Caillon-Caravanier, M.; Galiano, H.; Lemordant, D. Physicochemical Characterization of Morpholinium Cation Based Protic Ionic Liquids Used as Electrolytes. *Journal of Physical Chemistry B* 2010, *114*, 1757-1766.

(22)     Yue, C. B. Aromatic Compounds Mannich Reaction Using Economical Acidic Ionic Liquids Based on Morpholinium Salts as Dual Solvent-Catalysts. *Synthetic Communications* 2010, *40*, 3640-3647.
(23)     Zhang, Q. S.; Liu, A. X.; Guo, B. N.; Wu, F. Synthesis of Ionic Liquids Based on the N-Methyl-N-Allyl Morpholinium Cation. *Chemical Journal of Chinese Universities-Chinese* 2005, *26*, 340-342.
(24)     Lava, K.; Binnemans, K.; Cardinaels, T. Piperidinium, Piperazinium and Morpholinium Ionic Liquid Crystals. *Journal of Physical Chemistry B* 2009, *113*, 9506-9511.
(25)     Yeon, S. H.; Kim, K. S.; Choi, S.; Lee, H.; Kim, H. S.; Kim, H. Physical and Electrochemical Properties of 1-(2-Hydroxyethyl)-3-Methyl Imidazolium and N-(2-Hydroxyethyl)-N-Methyl Morpholinium Ionic Liquids. *Electrochimica Acta* 2005, *50*, 5399-5407.
(26)     Galinski, M.; Stepniak, I. Morpholinium-Based Ionic Liquid Mixtures as Electrolytes in Electrochemical Double Layer Capacitors. *Journal of Applied Electrochemistry* 2009, *39*, 1949-1953.
(27)     Domanska, U.; Lukoshko, E. V. Thermodynamics and Activity Coefficients at Infinite Dilution for Organic Solutes and Water in the Ionic Liquid 1-Butyl-1-Methylmorpholinium Tricyanomethanide. *Journal of Chemical Thermodynamics* 2014, *68*, 53-59.
(28)     Cizek, J. Origins of Coupled-cluster Technique for Atoms and Molecules. *Theoretica Chimica Acta* 1991, *80*, 91-94.
(29)     Pople, J. A.; Headgordon, M.; Raghavachari, K. Quadratic Configuration-Interaction - a General Technique for Determining Electron Correlation Energies. *Journal of Chemical Physics* 1987, *87*, 5968-5975.
(30)     Schmidt, M. W.; Baldridge, K. K.; Boatz, J. A.; Elbert, S. T.; Gordon, M. S.; Jensen, J. H.; Koseki, S.; Matsunaga, N.; Nguyen, K. A.; Su, S. J. et al. General Atomic and Molecular Electronic-Structure System. *Journal of Computational Chemistry* 1993, *14*, 1347-1363.
(31)     Chaban, V. V.; Prezhdo, O. V. Polarization Versus Temperature in Pyridinium Ionic Liquids. *Journal of Physical Chemistry B* 2014, *118*, 13940-13945.
(32)     Chaban, V. Competitive Solvation of the Imidazolium Cation by Water and Methanol. *Chemical Physics Letters* 2015, *623*, 76-81.